\documentclass[twocolumn,english,prl,floatfix,superscriptaddress,showpacs]{revtex4-1}

\usepackage[T1]{fontenc}
\usepackage{times}
\usepackage{color,graphicx}
\usepackage{subfig,array}
\usepackage{amsthm,amssymb,amsmath}
\usepackage{float}

\newcommand\ket[1]{\ensuremath{|#1\rangle}}

\newcommand{\nc}{\newcommand}
\nc{\cA}{{\cal A}} \nc{\cB}{{\cal B}} \nc{\cC}{{\cal C}}
\nc{\cD}{{\cal D}} \nc{\cE}{{\cal E}} \nc{\cF}{{\cal F}}
\nc{\cG}{{\cal G}} \nc{\cH}{{\cal H}} \nc{\cI}{{\cal I}}
\nc{\cJ}{{\cal J}} \nc{\cK}{{\cal K}} \nc{\cL}{{\cal L}}
\nc{\cM}{{\cal M}} \nc{\cN}{{\cal N}} \nc{\cO}{{\cal O}}
\nc{\cP}{{\cal P}} \nc{\cQ}{{\cal Q}} \nc{\cR}{{\cal R}}
\nc{\cS}{{\cal S}} \nc{\cT}{{\cal T}} \nc{\cU}{{\cal U}}
\nc{\cV}{{\cal V}} \nc{\cW}{{\cal W}} \nc{\cX}{{\cal X}}
\nc{\cZ}{{\cal Z}}

\begin{document}


\title{Topological and Error-Correcting Properties for
  Symmetry-Protected Topological Order}

\author{Bei Zeng}%
\affiliation{Department of Mathematics \& Statistics, University of
  Guelph, Guelph, Ontario, Canada}%
\affiliation{Institute for Quantum Computing, University of Waterloo,
  Waterloo, Ontario, Canada}%
\affiliation{Canadian Institute for Advanced Research, Toronto,
  Ontario, Canada}%
\author{Duan-Lu Zhou} %
\affiliation{Beijing National Laboratory for Condensed Matter Physics,
  and Institute of Physics, Chinese Academy of Sciences, Beijing
  100190, China}

\begin{abstract}
  We discuss the symmetry-protected topological (SPT) orders for
  bosonic systems from an information-theoretic viewpoint. We show
  that with a proper choice of the onsite basis, the degenerate
  ground-state space of SPT orders (on a manifold with boundary) is a
  quantum error-correcting code with macroscopic classical distance,
  hence is stable against any local bit-flip errors. We show that this
  error-correcting property of the SPT orders has a natural connection
  to that of the symmetry-breaking orders, whose degenerate
  ground-state space is a classical error-correcting code with a
  macroscopic distance, providing a new angle for the hidden
  symmetry-breaking properties in SPT orders. We propose new types of
  topological entanglement entropy that probe the STP orders hidden in
  their symmetric ground states, which also signal the topological
  phase transitions protected by symmetry. Combined with the original
  definition of topological entanglement entropy that probes the
  `intrinsic topological orders', and the recent proposed one that
  probes the symmetry-breaking orders, the set of different types of
  topological entanglement entropy may hence distinguish topological
  orders, SPT orders, and symmetry-breaking orders, which may be mixed up
  in a single system.
\end{abstract}

\date{\today}

\maketitle

\textit{Introduction} -- Symmetry protected topological (STP) orders
are gapped phases of matter with certain symmetry and only short-range
entanglement. It has been a focus of the recent studies in condensed
matter physics due to the excitement of the new experimental advances
in topological insulators and superconductors~\cite{QZ11}. The
classification of free fermionic STP phases are well
understood~\cite{kitaev2009periodic}. The situation of the interacting
systems are more complicated, with extensively recent discussions for
both the bosonic
case~\cite{gu2009tensor,chen2012symmetry,chen2013symmetry,kapustin2014symmetry}
and the fermionic
case~\cite{turner2011topological,fidkowski2011topological,gu2012symmetry}.

While many recent literatures are focusing on the symmetry aspects of
the STP orders, we would like to examine more details regarding the
topological properties of these systems from an information-theoretic
viewpoint. We start with the discussion of bosonic systems in one
spatial dimension (1D), where the gapped ground states of local
Hamiltonians are extensively
studied~\cite{hastings2007area,chen2011classification,landau2013polynomial}.

\begin{figure}[h]
  \centering
  \includegraphics[scale=0.2]{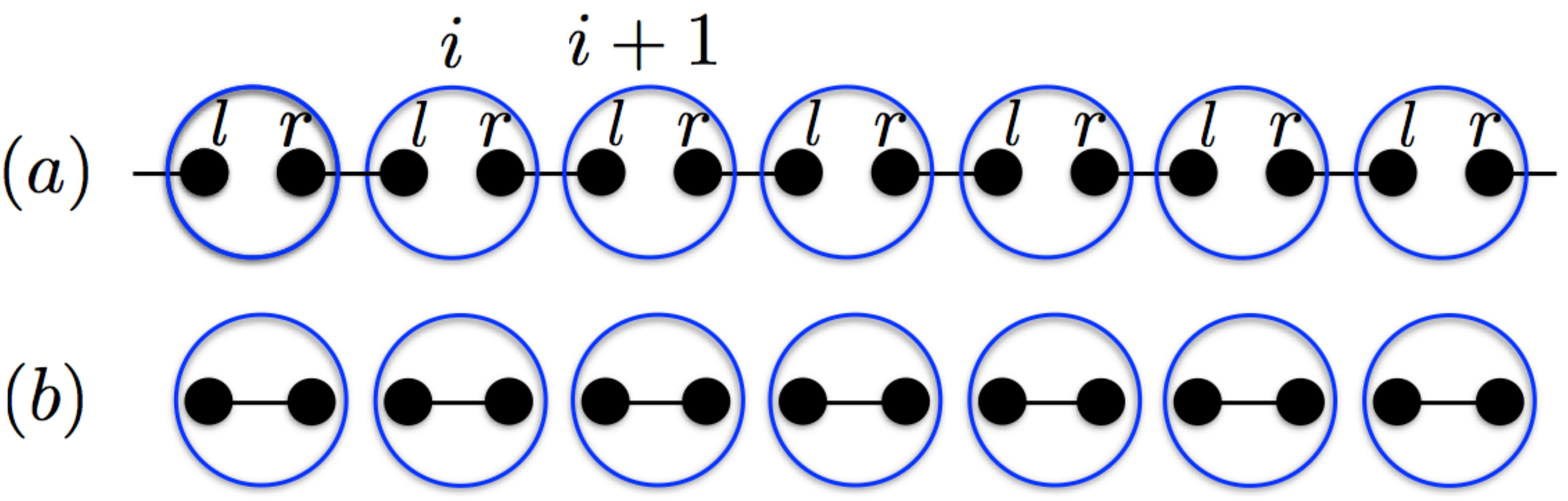}
  \caption{The MPS isometric form. The circles represent sites and the
    black dots represent (virtual) qubits (for simplicity we consider
    the qubit case, and our discussions naturally generalize to the
    qudit case). Two adjacent qubits connected by a line represent a
    bond, which is given by
    $\ket{w}=\frac{1}{\sqrt{2}}(\ket{00}+\ket{11})$. (a) Each site
    containing two qubits as label by $l$ (left) and $r$ (right). (b)
    Shifting the system by one (virtual) qubit.}
  \label{fig:SPT}
\end{figure}

It is well-known that for a 1D gapped Hamiltonian, the ground states
obey entanglement area
law~\cite{hastings2007area,chen2011classification,schuch2011classifying,arad2013area,landau2013polynomial,huang20141d}
and can be faithfully represented by the matrix product states
(MPS)~\cite{fannes1992finitely}. When the ground state is unique, the
MPS representation has injective matrices and can be adiabatically
connected to an isometric form (as shown in Fig.~\ref{fig:SPT}(a), for
periodic boundary conditions) via a renormalization procedure (with
possibly blocking of
sites)~\cite{chen2011classification,schuch2011classifying}.

Consider a system with $n$ sites hence total $2n$ (virtual) qubits,
and the quantum state of the system in Fig.~\ref{fig:SPT}(a) can be
written as $\ket{\Psi_a}=\otimes_i\ket{w}_{i_r,(i+1)_l}$, where the
label $i$ denotes sites, and the subscript $l$/$r$ of the site $i$
denotes the left/right (virtual) qubit in the site. If one further
applies a two-site unitary transformation on each bond, the system can be
disentangled to a product state $\ket{0}^{\otimes {2n}}$.

In order to reveal properly the nontrivial topological properties of
the system, certain symmetry is needed to prevent the system from
going to a trivial product state, which is the meaning of `symmetry
protection'. The distinct topological feature of an SPT state, for
instance the state $\ket{\Psi_a}$, is that when putting on a 1D chain
with boundary, each boundary carries an unpaired qubit, hence the
corresponding Hamiltonian has a $4$-fold degenerate ground state. This
is very different from the state
$\ket{\Psi_b}=\otimes_i\ket{w}_{i_l,{i}_r}$ as illustrated in
Fig.~\ref{fig:SPT}(b), by shifting $\ket{\Psi_a}$ by a (virtual)
qubit. $\ket{\Psi_b}$ essentially is a product state of onsite wave
functions, and does not carry any unpaired qubit on a 1D chain with
boundary.

$\ket{\Psi_b}$ clearly has the same symmetry as $\ket{\Psi_a}$.
However, when certain symmetry is respected (e.g.
$\mathbb{D}_2=\mathbb{Z}_2\times\mathbb{Z}_2$), $\ket{\Psi_a}$ cannot
be adiabatically connected to $\ket{\Psi_b}$ without a phase
transition. This phase transition is in this sense topological, which
however needs the symmetry protection to happen. It is shown that the
underlying reason for $\ket{\Psi_a}$ to be different from
$\ket{\Psi_b}$ is that they carry different projective representations
of the symmetry group, and theories based on the group cohomology may
be used to distinguish different SPT
phases~\cite{chen2011classification,schuch2011classifying}.

In this work, we propose a new approach to the theory of SPT orders
from an information-theoretic viewpoint. We show that, under a proper
choice of onsite basis, the degenerate ground-state space is a quantum
error-correcting code with a macroscopic classical distance, hence is
stable against any local bit-flip errors. This error-correcting
property has a natural connection to that of the symmetry-breaking
orders, whose degenerate ground-state space is a classical
error-correcting code with a macroscopic distance. Our approach hence
provides a new angle for the hidden symmetry-breaking properties in
SPT orders~\cite{PhysRevB.45.304,kennedy1992hidden,PBAO12,DQ13,EBD13}.

We further propose new types of topological entanglement entropy,
which probe the STP orders and signal the topological phase
transitions protected by symmetry. Our new types of topological
entanglement entropy are defined on a manifold with boundary, which
can probe the topological properties hidden in the symmetric ground states.

\textit{Error-Correcting Properties} -- To examine the
error-correcting properties for SPT orders, it is convenient to
transform the MPS isometric form into the cluster state model by an
onsite transformation. Notice that the bond state
$\ket{w}_{i_r,(i+1)_l}$ is a two-qubit stabilizer state with the
stabilizer generators $\{X_{i_r}X_{(i+1)_l},Z_{i_r}Z_{(i+1)_l}\}$. Now
on each site, we apply the transformation
\begin{equation}
  \label{eq:siteU}
  U_i=CNOT_{i_l,i_r}H_{i_r},
\end{equation}
where $CNOT_{i_l,i_r}$ is the controlled-NOT operation with the
$i_l$th qubit as the control qubit, and $H_{i_r}$ is the Hadamard
transformation on the $i_r$th qubit.

After the transformation $\prod_i U_i$, we have
\begin{eqnarray}
  X_{i_r}X_{(i+1)_l}&\rightarrow& Z_{i_r}X_{(i+1)_l}Z_{(i+1)_r}\nonumber\\
  Z_{i_r}Z_{(i+1)_l}&\rightarrow& Z_{i_l}X_{i_r}Z_{(i+1)_l},
\end{eqnarray}
which is the stabilizer generators for a 1D cluster state~\cite{BR01}
of $2n$ qubits.

Without loss of generality we now consider a 1D qubit system of $N$
qubits, with $N$ not necessarily even. And without confusion we label
each qubit by $j$. The 1D cluster state hence corresponds to the
stabilizer group with generators $\{Z_{j-1}X_jZ_{j+1}\}$, and the
corresponding Hamiltonian
\begin{equation}
  H_{clu}=-\sum_j Z_{j-1}X_jZ_{j+1}.
\end{equation}

For a 1D ring without boundary, the ground state of $H_{clu}$ is
unique. For a chain with boundary, where the summation index $j$ runs
from $2$ to $N-1$, the ground state is then $4$-fold degenerate. We
can also view the degenerate ground-state space as a quantum
error-correcting code encoding two qubits. As a quantum code, it has
only distance $1$, as $Z_1$ commutes with all the stabilizer
generators.

What we are interested in here is the ability of this code for
correcting classical errors (bit flip), which corresponds to errors
that are tensor products of $X_j$s. It is straightforward to see that
the two logical operators which are in the form of tensor products of
$X_j$s are
\begin{equation}
  \bar{X}_1=\prod_k X_{2k-1},\quad \bar{X}_2=\prod_k X_{2k},
\end{equation}
with $k$ runs from $1$ to $\lfloor N/2\rfloor$, and to interpret
$\prod_k X_{2k-1}$ containing also a product with $X_N$ if $N$ is odd.
This code hence has classical distance $\lfloor N/2\rfloor$, which is
a macroscopic distance that is half of the system size.

Another way to view $\bar{X}_1$ and $\bar{X}_2$ is that they generate
the group $\mathbb{D}_2=\mathbb{Z}_2\times\mathbb{Z}_2$ that preserves
the topological order of the system~\cite{else2012symmetry}. Any local perturbation respecting
the symmetry cannot lift the ground state degeneracy (in the
thermodynamical limit)~\cite{else2012symmetry,SB09,DB09}.

One way to view this symmetry protection is to add a magnetic field
along the $X$ direction to the system, and the corresponding
Hamiltonian reads
\begin{equation}
  H_{clu}(B)=-\sum_{j} Z_{j-1}X_jZ_{j+1}+B\sum_j X_j.
\end{equation}
It is known that there is a phase transition at $B=1$ (for periodic
boundary condition)~\cite{PP04,SB09,DB09}.

It is interesting to compare the system $H_{clu}(B)$ with a symmetry-breaking
ordered Hamiltonian
\begin{equation}
  H_{syb}(B)=-\sum_{j} Z_{j-1}Z_{j+1}+B\sum_j X_j,
\end{equation}
with the same symmetry $\mathbb{D}_2$ given by $\bar{X}_1$, $\bar{X}_2$.
The degenerate ground-state
space of $H_{syb}(0)$ is a classical error-correcting code with
distance $\lfloor N/2\rfloor$, and is spanned by
\begin{equation}
  \label{eq:basis}
  \ket{0000\ldots 00},\ket{0101\ldots 01},\ket{1010\ldots 10},\ket{1111\ldots 11}.
\end{equation}

Denote the symmetric ground state of $H_{syb}(B)$ by
$\ket{\psi_{syb}(B)}$. Then $\ket{\psi_{syb}(0)}$ is a stabilizer
state stabilized by $Z_{j-1}Z_{j+1}$ ($j=2,\ldots,N-2$) and
$\bar{X}_1$, $\bar{X}_2$, and is in fact a equal weight superposition
of the basis states of the code as given in Eq.~\eqref{eq:basis}.
Similarly, we denote the symmetric ground state of $H_{clu}(B)$ by
$\ket{\psi_{clu}(B)}$. Then $\ket{\psi_{clu}(0)}$ is a stabilizer
state stabilized by $Z_{j-1}X_jZ_{j+1}$ ($j=2,\ldots,N-2$) and
$\bar{X}_1$, $\bar{X}_2$.

There is no local unitary transformation to transform $H_{clu}(B)$ or
to $H_{syb}(B)$. One either needs a nonlocal transformation or a local
transformation with an unbounded depth, which reveals the hidden
symmetry-breaking property of the SPT order~\cite{DQ13,EBD13}. This
can also be seen from the fact that the symmetric ground
$\ket{\psi_{syb}(0)}$ is long-range entangled, and this long-range
property does not change even if closing the boundary. However, the
state $\ket{\psi_{clu}(0)}$, although appears to be long-range
entangled for a 1D chain with boundary (characterized by logical
operators $\bar{X}_1$, $\bar{X}_2$), is essentially short-range
entangled when closing the boundary. 

That is, $\ket{\psi_{clu}(0)}$ is
in fact stabilized by $Z_{j-1}X_jZ_{j+1}$ with a periodic boundary
condition~\cite{else2012symmetry}. 
In this sense, viewed as a dimension $0$ quantum code,
$\ket{\psi_{clu}(0)}$ also has a macroscopic classical
distance (given by the smallest weight element
in the stabilizer group which is a tensor product of $X_j$s~\cite{danielsen2005self}). 
Going along the direction respecting the symmetry picks up
the symmetric ground state as the exact ground state, which gives rise
to the phase transition for both the periodic and open boundary
conditions.

\textit{Topological Entanglement Entropy} -- Topological entanglement
entropy was first proposed to detect topological
orders~\cite{levin2006detecting,KP06}, and is recently generalized to
probe the systems with symmetry-breaking orders~\cite{MaxEnt,LIT}.
Inspired by these previous types of topological entanglement entropy,
we introduce new types to probe the SPT orders.

\begin{figure}[h]
  \centering
  \includegraphics[scale=0.3]{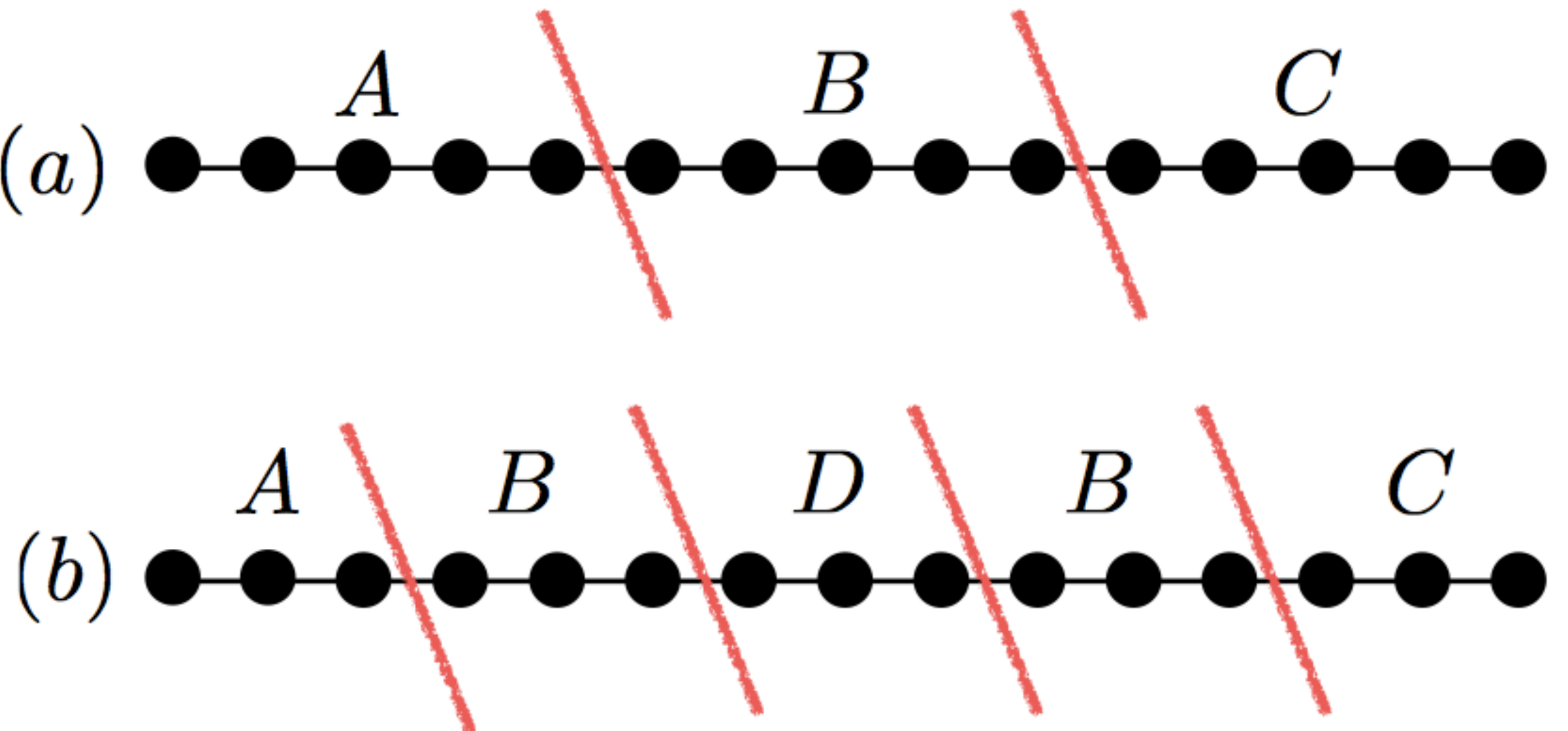}
  \caption{(a) Cutting a 1D chain into $A,B,C$ parts; (b) Cutting a 1D
    chain into $A,B,C,D$ parts.}
  \label{fig:cutting}
\end{figure}

We consider a 1D chain with boundary. For any gapped ground state, and
for the cuttings given in Fig.~\ref{fig:cutting}, we introduce the
topological entanglement entropy
\begin{equation}
  \label{eq:Stopo}
  S_{topo}=S_{AB}+S_{BC}-S_B-S_{ABC},
\end{equation}
where $S(*)$ is the von Neumann entropy of reduced density matrix of
the part $*$.

There are two kinds of cuttings introduced in Fig.~\ref{fig:cutting}.
Fig.~\ref{fig:cutting}$(a)$ cuts the system into three parts, and we
denote the corresponding topological entanglement entropy by
$S_{topo}^{t}$. Fig.~\ref{fig:cutting}$(b)$ cuts the system into four
parts, and we denote the corresponding topological entanglement
entropy by $S_{topo}^{q}$. We use $S_{topo}$ to refer both
$S_{topo}^{t}$ and $S_{topo}^{q}$.

Similarly to the topological entanglement entropy introduced
previously, $S_{topo}$ is an invariant of local unitary
transformations and $S_{topo}=0$ for unique gapped ground
states~\cite{levin2006detecting,KP06,LIT}. We also know that
$S_{topo}$ is quantized for SPT ordered states due to their degenerate
entanglement spectrum~\cite{PhysRevLett.101.010504}, hence a nonzero
$S_{topo}$ is a signature of SPT order. We will show that $S_{topo}$
also signals topological phase transitions protected by symmetry.

\begin{figure}[h]
  \centering
  \includegraphics[scale=0.24]{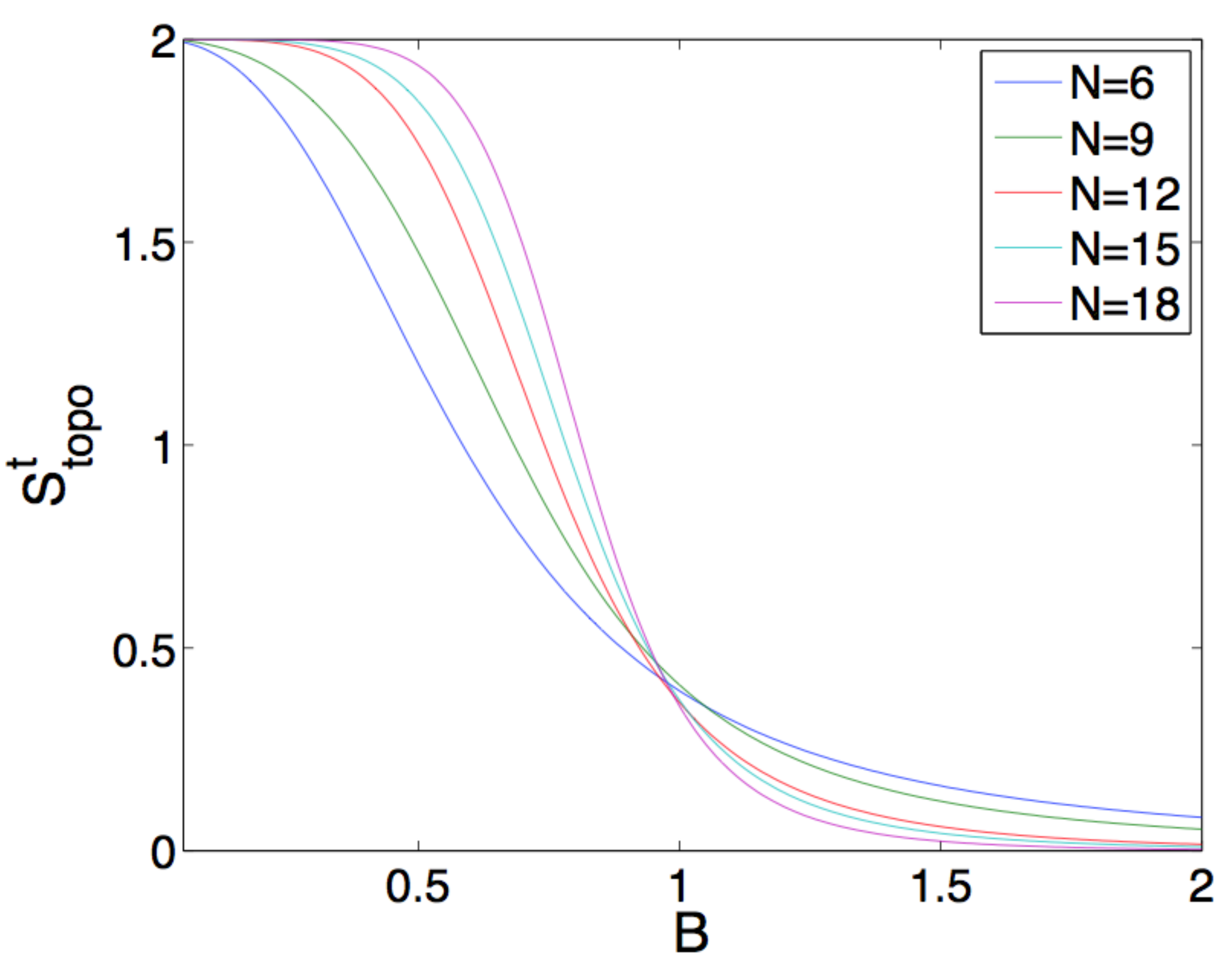}
  \caption{$S^{t}_{topo}$ for the ground state of $H_{clu}$.}
  \label{fig:cluster}
\end{figure}

We first examine $S^{t}_{topo}$. For the ideal state of $B=0$,
$S^{t}_{topo}=2$ for both $\ket{\psi_{clu}(0)}$ and
$\ket{\psi_{syb}(0)}$. When $B$ increases, for $\ket{\psi_{clu}(B)}$,
$S^{t}_{topo}$ signals the topological phase transition, as shown in
Fig.~\ref{fig:cluster}.
However, the symmetry-breaking order hidden in the exact symmetric
ground state $\ket{\psi_{syb}(B)}$ can also be detected by
$S^{t}_{topo}$. In fact, for the same calculation with $6,9,12,15,18$
qubits, one gets a very similar figure as Fig.~\ref{fig:cluster}.

To distinguish SPT orders from a symmetry-breaking one, we can instead
use $S^{q}_{topo}$. Since the topological entanglement entropy is only
carried in the entire wave function of the exact symmetric ground
state for symmetry-breaking orders~\cite{LIT}, computing
$S^{q}_{topo}$ on its reduced density matrix of parts $ABC$ returns
$0$.

\begin{figure}[h]
  \centering
  \includegraphics[scale=0.24]{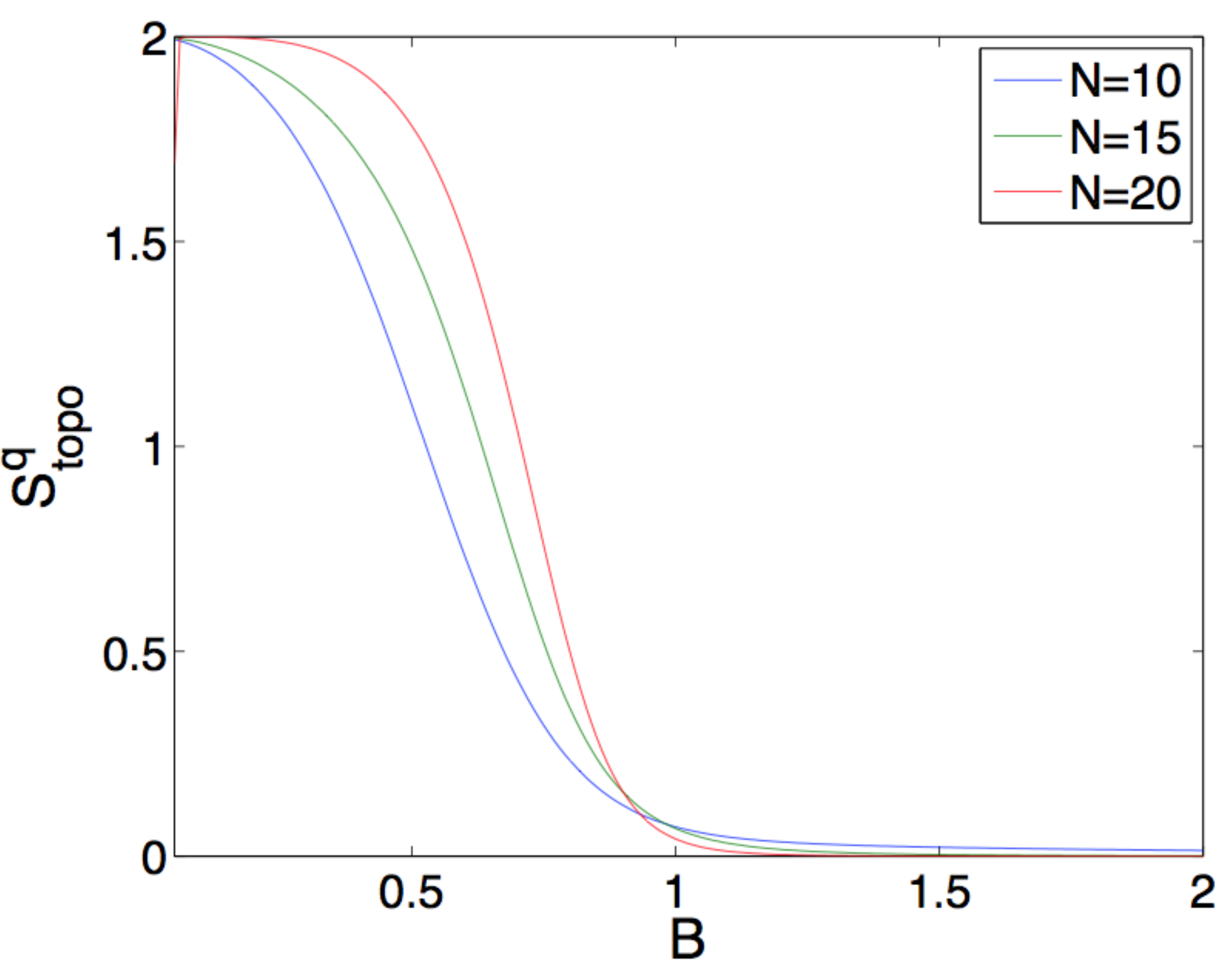}
  \caption{$S^{q}_{topo}$ for the ground state of $H_{clu}$.}
  \label{fig:clusterD}
\end{figure}

However, $S^{q}_{topo}=2$ for $\ket{\psi_{clu}(0)}$, because the
`topology' of the STP states is essentially carried on the boundary,
tracing out part of the bulk has no effect on detecting the
topological order. For $\ket{\psi_{clu}(B)}$, $S^{q}_{topo}$ signals
the topological phase transition, as shown in Fig.~\ref{fig:clusterD}.

\textit{A mixing order of symmetry-breaking and SPT} -- There could be
also systems containing mixing orders of symmetry-breaking and SPT,
whose symmetric ground states correspond to non-injective matrices in the MPS
representations, with isometric forms that couple GHZ states
with short-ranged bond states~\cite{schuch2011classifying}.

As an example, we consider a stabilizer group generated by
$Z_{j-1}X_{j}X_{j+1}Z_{j+2}$ with $j$ running from $2$ to $N-2$, which is
a generalization of the $5$-qubit code~\cite{LMPZ96,BDSW96} and a
special kind of quantum convolutional
codes~\cite{grassl2007constructions}. On a 1D chain with boundary,
i.e. for $j=2,3,\ldots, N-2$, the Hamiltonian $-\sum_j
Z_{j-1}X_{j}X_{j+1}Z_{j+2}$ has $8$-fold ground-state degeneracy.

The ground-state as an error-correcting code has classical distance
$\lfloor N/3\rfloor$, with logical operators $\bar{X}_1=\prod_k
X_{3k-2}$, $\bar{X}_2=\prod_k X_{3k-2}$, $\bar{X}_3=\prod_k X_{3k}$.
Therefore, if one adds a magnetic field along the $X$ direction, i.e.
\begin{equation}
  H_{ZXXZ}(B)=-\sum_j Z_{j-1}X_{j}X_{j+1}Z_{j+2}+B\sum_j X_j,
\end{equation}
the orders of the system (either SPT or symmetry-breaking) will be
protected.

\begin{figure}[h]
  \centering
  \includegraphics[scale=0.24]{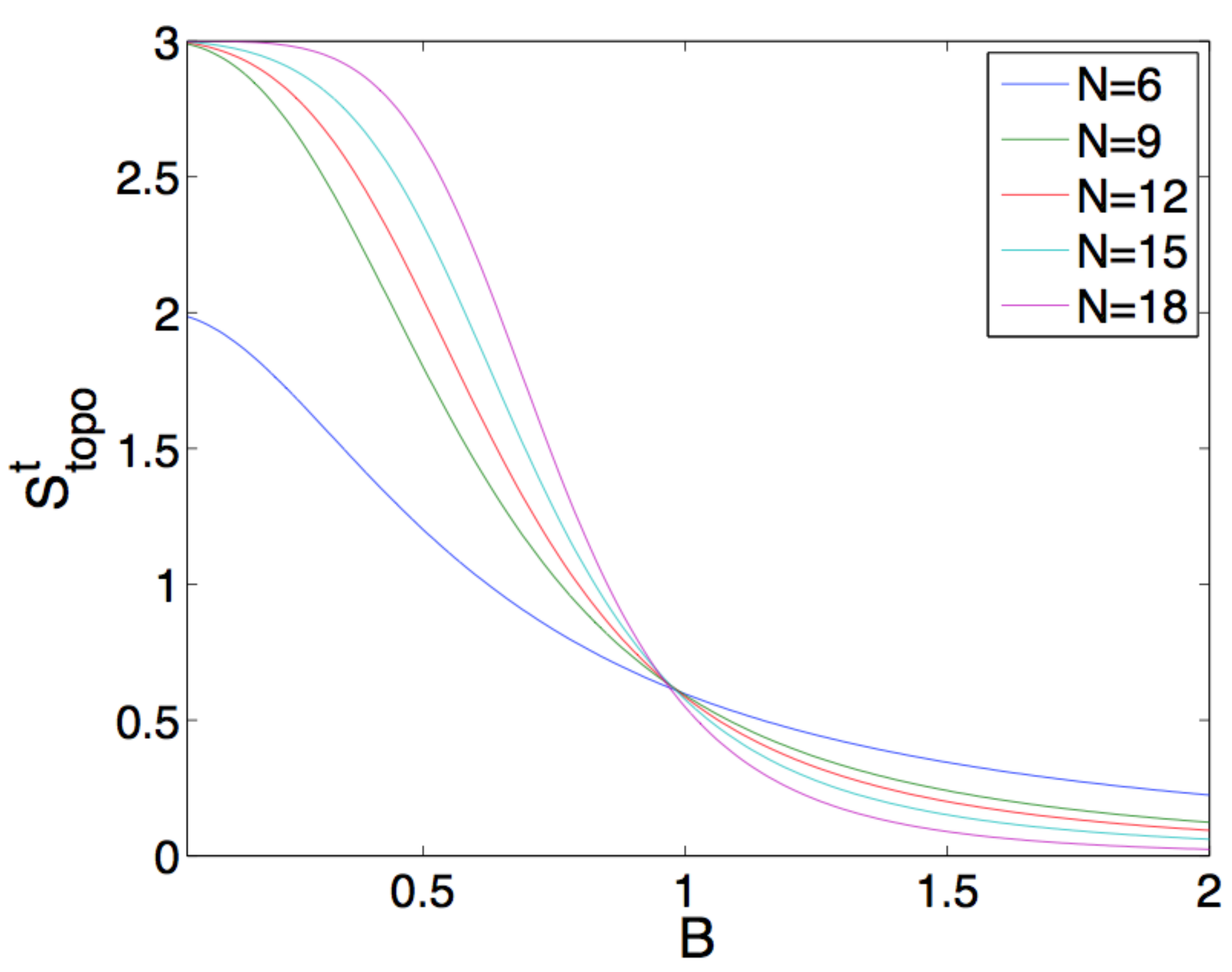}
  \caption{$S^{t}_{topo}$ the ground state of $H_{ZXXZ}(B)$. For
    $N=6$, the maximum value of $S^{t}_{topo}$ is $2$ due to that the
    system size is too small.}
  \label{fig:ZXXZ1}
\end{figure}

\begin{figure}[h]
  \centering
  \includegraphics[scale=0.24]{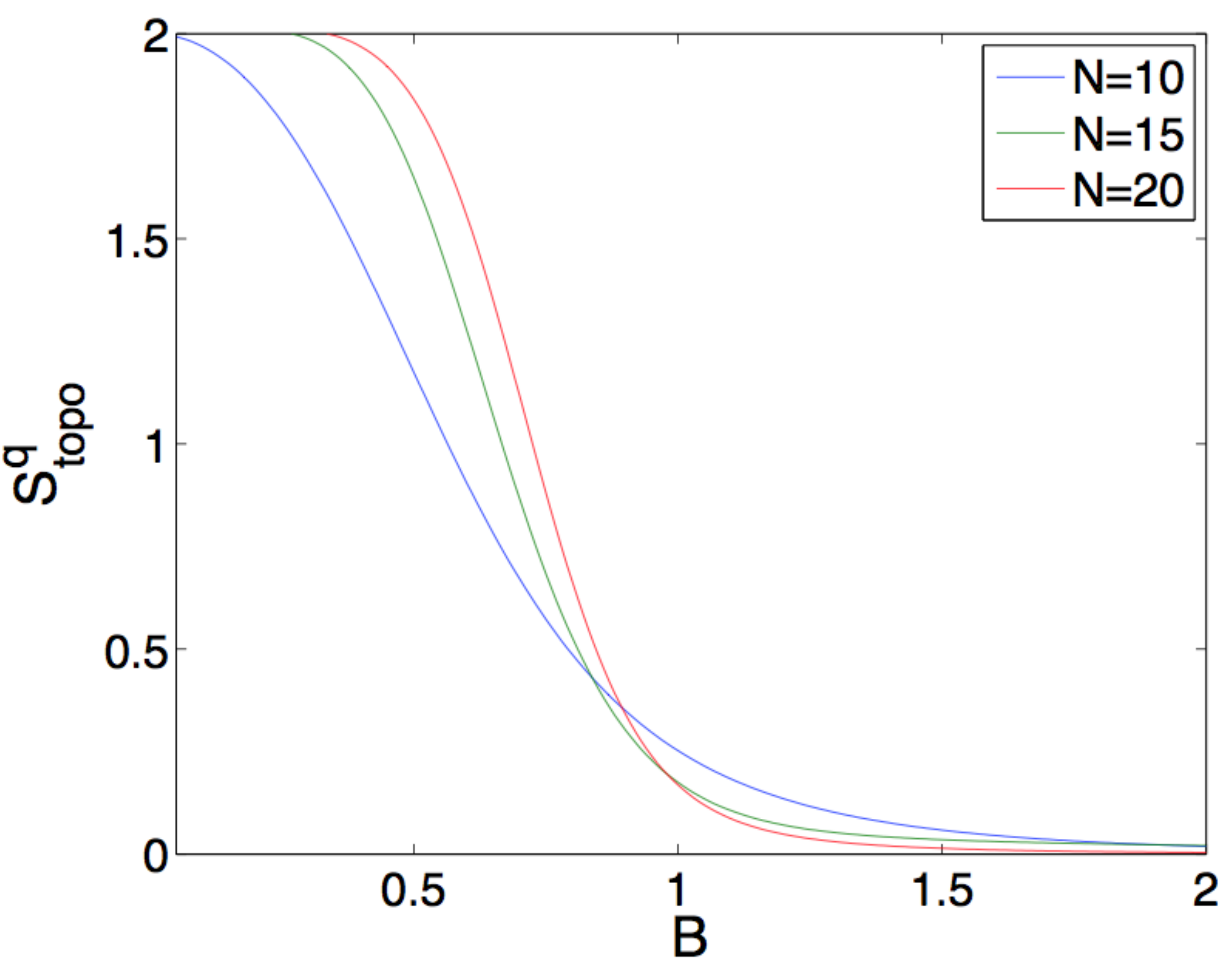}
  \caption{$S^{q}_{topo}$ for the ground state of $H_{ZXXZ}(B)$.}
  \label{fig:ZXXZ2}
\end{figure}

It turns out that the system combines a $\mathbb{Z}_2$
symmetry-breaking order and a $\mathbb{D}_2$ SPT-order. This can be
seen from the fact that for $B=0$, the symmetric ground state has
$S^{t}_{topo}=3$ and $S^q_{topo}=2$. $S^{t}_{topo}$ probes both the
symmetry-breaking order and the SPT order, as illustrated in
Fig.~\ref{fig:ZXXZ1}. $S^{t}_{topo}$ probes only the SPT phase
transition, as illustrated in Fig.~\ref{fig:ZXXZ2}.

\textit{Discussion} -- What Eq.~\eqref{eq:siteU} essentially does, is
to map the onsite state $\ket{w}_{i_l,i_r}$ illustrated in
Fig.~\ref{fig:SPT}(b) to a product state of qubits
$\ket{+}_{i_l}\otimes\ket{+}_{i_r}$. Here
$\ket{+}=\frac{1}{\sqrt{2}}(\ket{0}+\ket{1})$ is the eigenvalue $1$
eigenstate of $X$. Because going from the state $\ket{\Psi_a}$ (the
state illustrated in Fig.~\ref{fig:SPT}(a)) to $\ket{\Psi_b}$ while
respecting symmetry will encounter a phase transition, directly
interpolating the the cluster state to $\ket{+}^{\otimes 2n}$ (i.e.
given by $H_{clu}(B)$) also undergoes a phase transition. Therefore,
the onsite transformation $\prod_i U_i$ transforms $\ket{\Psi_b}$ to
the symmetric ground state of a quantum error-correcting code with a
macroscopic classical distance.

This idea can be generalized to higher spatial dimensions. In a
general setting, an SPT ordered state $\ket{\Phi_a}$ is that, when
connecting to a product state $\ket{\Phi_b}$ with the same symmetry, a
phase transition occurs while respecting the symmetry~\cite{CGZW13}.
One can always apply some onsite unitary transformation to transform
$\ket{\Phi_b}$ to a tensor product of $\ket{+}$, hence at the same
time transform $\ket{\Phi_a}$ to the symmetric ground state of some
quantum error-correcting code with a macroscopic classical distance (for 
instance the SPT ordered 2D cluster state discussed in~\cite{else2012symmetry}).

One may also generalize the idea of different types of topological
entanglement entropy to higher spatial dimensions. 
For instance, in 2D, a straightforward way is to
replace the chain by a cylinder with boundary, then use the 
similar cuttings as in Fig.~\ref{fig:cutting}.

\begin{figure}[h]
  \centering
  \includegraphics[scale=0.25]{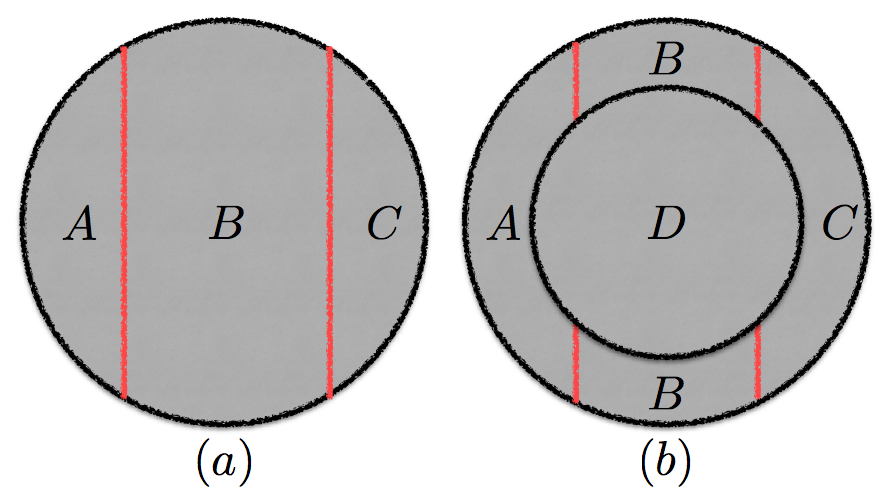}
  \caption{(a) Cutting a 2D disk into $A,B,C$ parts; (b) Cutting a 2D
    disk into $A,B,C,D$ parts.}
  \label{fig:cutting2D}
\end{figure}

One may also consider a disk with boundary. For any gapped ground state
(one may need to avoid the situation of a gapless boundary by adding symmetric local
 terms to the Hamiltonian), still
using $S_{topo}$ as given in Eq.~\eqref{eq:Stopo}, one can consider
two kinds of cuttings, as given in Fig.~\ref{fig:cutting2D}.
Similarly as the 1D case, the cutting of Fig.~\ref{fig:cutting2D}(a)
probes both the symmetry-breaking orders and the STP orders, and the
cutting of Fig.~\ref{fig:cutting2D}(a) probes only STP orders.

Notice that the topological entanglement entropy proposed
in~\cite{MaxEnt,LIT} is defined on a manifold without boundary (e.g. a
1D ring or a 2D sphere), which detects only symmetry-breaking orders
but not SPT orders. Combined with the original definition of
topological entanglement entropy~\cite{levin2006detecting,KP06} that
probes the `intrinsic topological orders', and the recent proposed one
that probes the symmetry-breaking orders~\cite{MaxEnt,LIT}, the set of
different types of topological entanglement entropy may hence
distinguish topological orders, SPT orders, and symmetry-breaking
orders, which may be mixed up in a single system.

We hope our discussion adds new ingredients for understanding the
microscopic theory of SPT orders.

\textit{Acknowledgements} -- We thank Stephen Bartlett for explaining
the hidden symmetry-breaking in SPT orders. We thank Xingshan Cui,
Eric Rowell, Zhenghan Wang, and Xiao-Gang Wen for helpful discussions
on STP orders. We thank Markus Grassl and Martin Roetteler for
helpful discussions on quantum convolutional codes. We are grateful to Yang
Liu for providing codes for exact diagonalization. BZ is supported by
NSERC. DLZ is supported by NSF of China under Grant No. 11175247, and
NKBRSF of China under Grants Nos. 2012CB922104 and 2014CB921202.

\bibliography{TopEnt}

\end{document}